\documentclass[prl,twocolumn,showpacs,superscriptaddress,letterpaper]{revtex4-1}

\usepackage{graphicx}

\begin{document}

\title{Soft antiphase tilt of oxygen octahedra in the hybrid improper
    multiferroic $\rm Ca_3Mn_{1.9}Ti_{0.1}O_7$}

\author{Feng Ye}
\email{yef1@ornl.gov}
\affiliation{Quantum Condensed Matter Division, Oak Ridge National Laboratory, Oak Ridge, Tennessee 37831, USA}
\author{Jinchen Wang}
\author{Jieming Sheng}
\affiliation{Department of Physics, Renmin University of China, Beijing 100872, China}
\affiliation{Quantum Condensed Matter Division, Oak Ridge National Laboratory, Oak Ridge, Tennessee 37831, USA}
\author{C.~Hoffmann}
\affiliation{Chemical and Engineering Materials Division, Oak Ridge National Laboratory, Oak Ridge, Tennessee 37831, USA}
\author{T.~Gu}
\author{H.~J.~Xiang}
\affiliation{Key Laboratory of Computational Physical Sciences, State Key
    Laboratory of Surface Physics, and Department of Physics, Fudan
    University, Shanghai 200433, China }
\affiliation{Collaborative Innovation Center of Advanced Microstructures, Nanjing 210093, China}
\author{Wei~Tian}
\affiliation{Quantum Condensed Matter Division, Oak Ridge National Laboratory, Oak Ridge, Tennessee 37831, USA}
\author{J.~J.~Molaison}
\affiliation{Instrument and Source Division, Oak Ridge National Laboratory, Oak Ridge, Tennessee 37831, USA}
\author{A.~M.~dos~Santos}
\author{M.~Matsuda}
\author{B.~C.~Chakoumakos}
\affiliation{Quantum Condensed Matter Division, Oak Ridge National Laboratory, Oak Ridge, Tennessee 37831, USA}
\author{J.~A.~Fernandez-Baca}
\affiliation{Quantum Condensed Matter Division, Oak Ridge National Laboratory, Oak Ridge, Tennessee 37831, USA}
\author{X.~Tong}
\affiliation{Instrument and Source Division, Oak Ridge National Laboratory, Oak Ridge, Tennessee 37831, USA}
\author{Bin Gao}
\author{Jae Wook Kim}
\author{S.-W. Cheong}
\affiliation{ Rutgers Center for Emergent Materials and Department of Physics and Astronomy,
Rutgers University, Piscataway, New Jersey 08854, USA}
\date{\today}

\begin{abstract}
We report a single crystal neutron and x-ray diffraction study of
the hybrid improper multiferroic $\rm Ca_3Mn_{1.9}Ti_{0.1}O_7$ (CMTO), a prototypical
system where the electric polarization arises from the condensation of two
lattice distortion modes. With increasing temperature ($T$), the
out-of-plane, antiphase tilt of MnO$_6$ decreases in amplitude
while the in-plane, inphase rotation remains robust and experiences abrupt
changes across the first-order structural transition.  Application of
hydrostatic pressure ($P$) to CMTO at room temperature shows a similar
effect. The consistent behavior under both $T$ and $P$ reveals the
softness of antiphase tilt and highlights the role of the partially occupied
$d$ orbital of the transition metal ions in determining the stability of
the octahedral distortion. Polarized neutron analysis indicates the
symmetry-allowed canted ferromagnetic moment is less than 0.04 $\mu_B$/Mn
site, despite a substantial out-of-plane tilt of the MnO$_6$ octahedra.
\end{abstract}


\maketitle

Multiferroic compounds with spontaneous elastic, electrical,
magnetic orders are considered as the key materials to achieve cross-control
between magnetism and electricity in solids with small energy
dissipation \cite{Cheong07,Tokura14}. The functional properties including
colossal magnetoelectric effect could be used in solid-state memories and
sensors \cite{Lee13}. The desired multifunctional behavior requires common
microscopic origin of the long-range order such that one order parameter is
strongly coupled to the conjugate field of the other one.  So far, the
majority of attention has focused on exploring materials with magnetic origin,
where the underlying microscopic mechanisms are primarily classified into
three types: symmetric spin exchange interaction $\Sigma_{ij}({\bf S}_i\cdot
{\bf S}_j$) \cite{Hur04,Xiang11}, antisymmetric spin-exchange interaction
${\bf S}_i\times {\bf S}_j$ \cite{Katsura05,Mostovoy06}, and spin-dependent
$p-d$ hybridization due to spin-ligand interaction  $({\bf e}_{il} \cdot {\bf
S}_{i})^2{\bf e}_{il}$ \cite{Arima07}. The material-by-design efforts focusing
on magnetic oxides has been productive; the ferroelectricity is induced either
through epitaxial strain engineering or chemical substitution of
stereochemical inactive ions with lone-pair-active cations
\cite{Wang03,Catalan05}.

However, this approach requires a strong coupling between ferroelectricity and
magnetism. The microscopic mechanism with spin origin also implicitly suggests
a low operating temperature because of the magnetic frustration.  On the other
hand, perovskites in the form of $ABO_3$ and their derivatives are favorably
chosen for functional materials due to their high susceptibility toward polar
structural instability and the intimate coupling between the ferroelectric
polarization and the magnetic, orbital, and electronic degrees of freedom.
Recently, a novel mechanism termed as {\it ``hybrid improper ferroelectric''} has
been proposed to search for materials with spontaneous ferroelectricity.  The
central idea is that the polar mode is driven by the condensation of two
nonpolar lattice modes, which represent oxygen octahedral rotation ($X^{+}_2$)
and tilt ($X^{-}_3$), respectively \cite{Benedek11,Benedek12,Benedek15}. It
is shown that ferroelectricity can be induced experimentally via strain
coupling of octahedral rotation between different perovskite heterostructure
layers \cite{Bousquet08} and even realized in bulk
materials \cite{Benedek11,Harris11}. Indeed, the higher-order coupling of
multidegrees of freedom with the polarization is observed in the
Ruddlesden-Popper structures $\rm Ca_3Ti_2O_7$ (CTO) and $\rm Ca_3Mn_2O_7$
(CMO) by neutron powder diffraction \cite{Senn15}. Both compounds exhibit the
low-temperature ($T$) polar $A2_1am$ space group (SG No.~36) that corresponds
to symmetry space spanned by two centrosymmetric space groups $Acam$ (No.~64)
and $Amam$ (No.~63). In the magnetically ordered CMO, the antiphase octahedral
tilt is responsible for the canted moment and suggests a controllable
polarization-magnetization coupling via modification of the oxygen octahedral
distortion \cite{Benedek11,Harris11}. Thus, instead of searching for materials
where the magnetism couples to the electric polarization, the proposal of
octahedral-distortion-driven ferroelectricity and magnetoelectricity in this
system has shifted the emphasis to discover room-temperature antiferromagnet
$A_3B_2O_7$ compounds \cite{Pitcher15,Oh15}. A deeper understanding of the
origin for octahedral distortion would provide valuable insight to design
materials with optimal performance.

\begin{figure}[thb!]
    \includegraphics[width=3.2in]{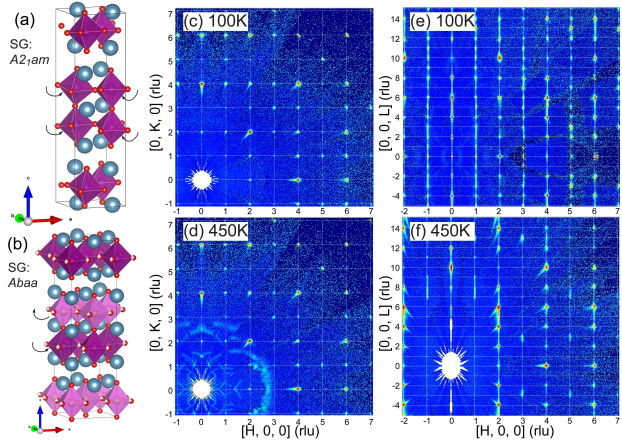}
    \caption{ Crystal structure of $\rm Ca_3Mn_{1.9}Ti_{0.1}O_7$ in (a) the
    low-$T$ $A2_1am$ and (b) high-$T$ $Abaa$ phases. The structures are
    drawn using VESTA software \cite{Momma11}. The diffraction images
    in the $(h,k,0)$ scattering plane at (c) 100~K and (d) 450 K. Images in
    the $(h,0,l)$ plane at (e) 100 K and (f) 450 K.
	}
\end{figure}

In this Rapid Communication, we report an active control of the octahedral distortion of a
prototypical hybrid improper multiferroic $\rm Ca_3Mn_{1.9}Ti_{0.1}O_7$ (CMTO)
using temperature ($T$) and pressure ($P$). Introducing small amounts of Ti
stabilized the crystal growth while keeping the physical properties 
similar to the pure CMO. We found the antiphase tilt of MnO$_6$ decreases in
amplitude while the inphase rotation remains nearly unchanged with increasing
temperature. Hydrostatic pressure has a similar effect; it suppresses the antiphase tilt
and has minimal influence on the inphase rotation. The consistent behavior of
two distortion modes with respect to both $T$ and $P$ reveals the mechanism
contributing to the distortion instability and indicates the ferroelectricity
of CMTO can be controlled indirectly by the amplitude of the octahedral distortion.

Single crystals of CMTO were grown using the traveling floating zone method.
The chemical compositions were determined by x-ray and neutron diffraction
refinements independently and physical properties were characterized by
resistivity, and magnetization measurements.  Single crystal x-ray diffraction
data were collected using a Rigaku XtaLAB PRO diffractometer at the Oak Ridge
National Laboratory.  Extensive neutron scattering measurements were
performed using instruments at the Spallation Neutron Source (SNS) and the
High Flux Isotope Reactor (HFIR).  Structure analysis was performed using the
TOPAZ diffractometer at SNS with crystal size of $\rm 2.5 \times 2.5\times
0.25~mm^3$. Large pieces were used to study the thermal evolution
of specific reflections using the single crystal diffuse scattering
spectrometer CORELLI at SNS, and the HB1A triple-axis spectrometer at HFIR.
Polarized neutron diffraction was performed on a sample with mass $\sim$
200~mg using the HB1 triple axis spectrometer. Heusler crystals were used as
the monochromator and analyzer for the polarization setup with flipping ratio
of 15 at incident energy $E_i=13.5$~meV.  High pressure neutron
diffraction was carried out at the SNAP instrument at SNS.  An oriented
crystal was loaded inside the Paris-Edinburgh pressure cell with lead powder
serving as the pressure transmitting medium and gauge.

\begin{figure}[thb!]
\vskip +.3cm
\includegraphics[width=3.2in]{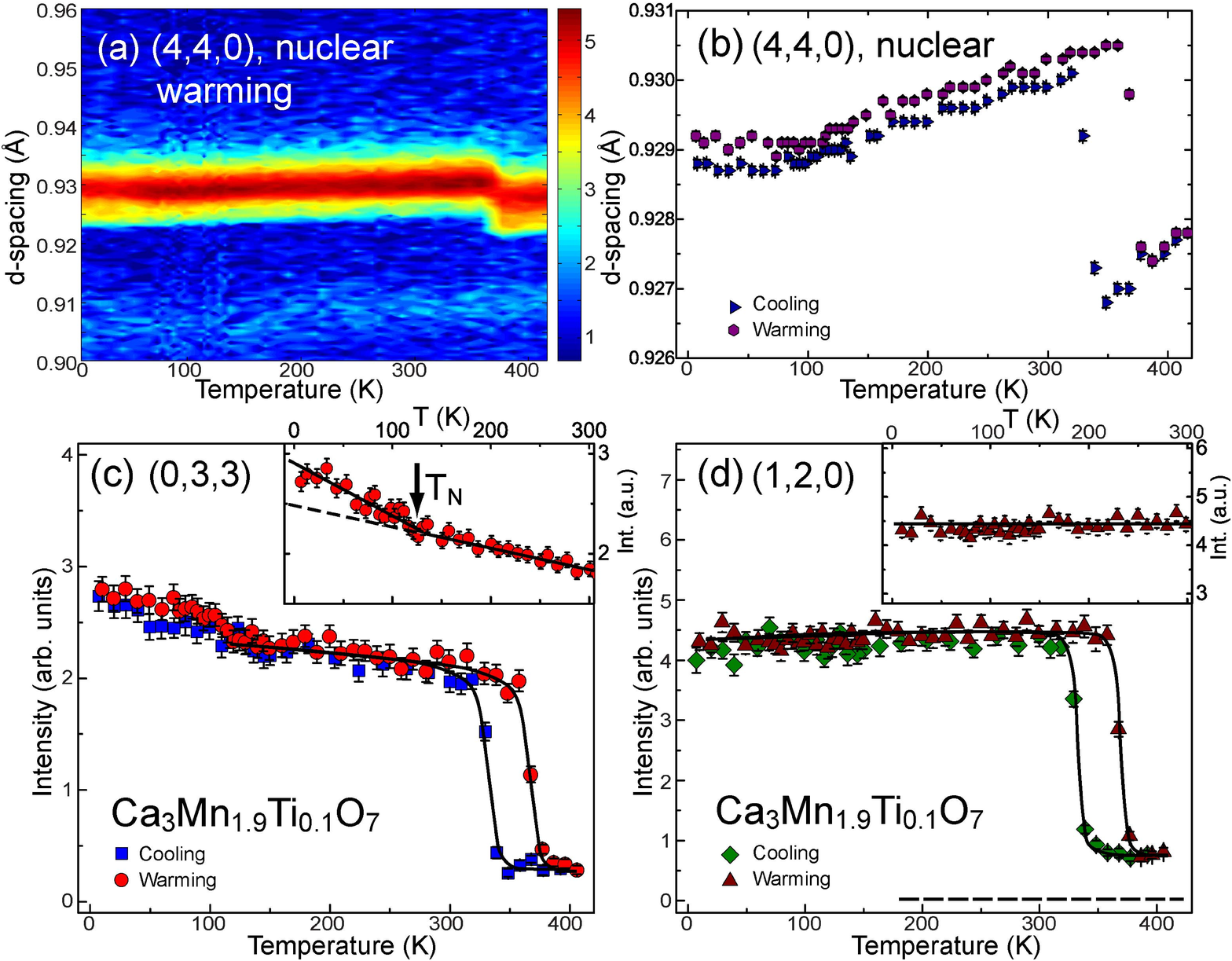}
\vskip -.2cm
    \caption{ (a) The $T$ dependence of the nuclear (4,4,0) reflection.
    (b) Comparison of the $d$ spacing of the same peak on cooling
    and warming. (c) The $T$ dependence of (0,3,3) reflection associated with
    the out-of-plane octahedral tilt.  (d) The $T$ dependence of the (1,2,0)
    peak from the in-plane rotation.  Insets of panels (c) and (d) show the
    zoom-in part near the antiferromagnetic transition.}
\end{figure}

Although the CMO crystal is known to have a non-centrosymmetric $A2_1am$
structure below room temperature \cite{Guiblin02,Lobanov04}, the evolution
pathway from the high-$T$ tetragonal to the low-$T$ orthorhombic phase remains
questionable since a direct continuous transition is not consistent with the
Landau theory. It was proposed that a certain intermediate phase might exist
over a narrow temperature range.  A synchrotron x-ray and neutron
diffraction study on the powder sample determined the structure above the
transition is nonpolar $Abaa$ (No.~68), where the $\rm MnO_6$ octahedra of
the neighboring perovskite block rotate out of phase by the same
angle \cite{Senn15}. The subtle structural transition and lack of sensitivity
in powder measurements motivates us to carry out a more thorough examination via
single crystal neutron diffraction. Figures 1(c)-1(f) compare the reciprocal
space images of the single crystal CMTO at 100 and 450~K in various
scattering planes. The low-$T$ patterns agree with the reflection condition
for the SG $A2_1am$ with Bragg reflections appearing at $k=2n$ in the
$(h,k,0)$ plane and $h+l=2n$ in the $(h,0,l)$ plane \cite{Ca327twinning}.  At
450~K, only reflections with $h=2n,l=2n$ appear in the $(h,0,l)$ plane, while
the pattern of the $(h,k,0)$ plane remains identical as compared to 100~K. The
existence of Bragg peaks with $h+k=\rm odd$ in the $(h,k,0)$ plane violates
the reflection condition for the nonpolar $Abaa$ phase. CMO has an abnormally
large temperature window ($\sim 70$~K) where the low-$T$ polar $A2_1am$ phase
coexists with the high-$T$ phase \cite{Senn15}. It is possible that the
crystal remains in the mixed-phase state even at 450~K.  Alternatively, the
observed reflection condition suggests a different crystal structure if
single phase is assumed. An exhaustive search leads to a polar $Aba2$ SG
(No.~41) which allows the $\rm MnO_6$ in the bilayer block to rotate
out-of-phase with different angles.  Refinement using such a structural model
improves the agreement between the model and data. Full details are given in
the Supplemental Material \cite{Ca327SM}.  Nevertheless, a definitive description of the
crystal structure above the transition requires complementary experimental
techniques, e.g., bulk polarization and imaging techniques like scanning
electron microscopy at high temperatures.

\begin{figure}[thb!]
\includegraphics[width=3.2in]{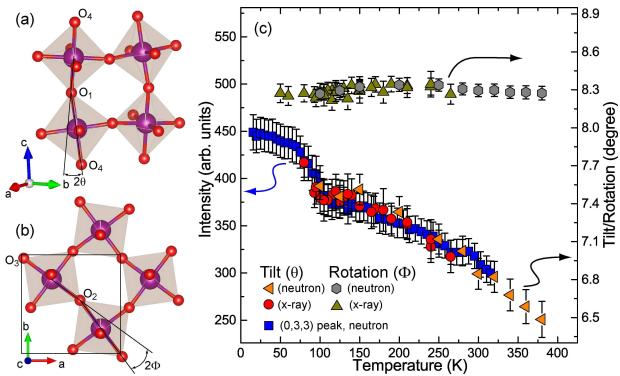}
    \caption{ The schematic diagram of (a) out-of-plane tilt and (b) in-plane
rotation of the MnO$_6$. (c) The $T$ dependence of refined $\rm MnO_6$
distortion angles. The $T$ dependence of the (0,3,3) peak intensity from
another independent measurement is plotted for comparison.}
\end{figure}

$T$ dependence of superlattice reflections is used to track the evolution
of the structural distortion. As shown in Figs.~2(a) and 2(b), a first-order
structural transition is evident near $\rm T_S\sim$370~K, where the lattice
constant experiences an abrupt change with a strong hysteresis.  The low-$T$
noncentrosymmetric phase is comprised of an in-plane rotation of $\rm MnO_6$
octahedra along [0,0,1] (i.e., $a^0a^0c^+$ in Glazer's
notation \cite{Glazer72})  and a out-of-plane tilt along the [1,1,0] axis in
the pseudo-tetragonal setting ($a^-a^-c^0$). We select two characteristic
vectors (1,2,0) and (0,3,3) where the scattering structure factors arises from
distinct instability modes.  Figure 2(c) shows that the peak intensity of
the tilt-active (0,3,3) reflection is clearly $T$-dependent, and decreases
steadily with increasing temperature. A change of slope near $T_N \sim 115$~K
is discernible [inset of Fig. 2(c)].  In contrast, the intensity of
the rotation-active (1,2,0) reflection remains flat below the structural phase
transition.  The stark difference indicates the rotation of $\rm MnO_6$ is
robust while the tilt is more susceptible to the thermal fluctuation.

The distortions depicted in Figs.~3(a) and 3(b) are quantified by the structural
refinement from both x-ray and neutron diffraction [different symbols in
Fig.~3(c)].  The inplane rotation and out-of-plane tilt are characterized by
the deviation of oxygen sites away from their undistorted positions. The
corresponding angles $\phi$ and $\theta$ can be estimated from the refined
coordinates, e.g., the rotation angle  $\phi= (u_1-u_2+v_1+v_2)$ for the
$\rm O_2$ and $\rm O_3$ sites located at $(u_1+1/2,v_1+1/2,w_1)$ and
$(u_2,1-v_2,w_2)$ [Fig.~3(b)].  The average $\phi$ is 8.3$^\circ$ in the basal
plane, larger than $\theta$ which is in the range of 6.5-7.7$^\circ$. Most
importantly, the refinement provides compelling evidence that the tilt
decreases progressively on warming, while the rotation remains rigid across
the whole temperature range.  This is in excellent agreement with the
diffraction data shown in Figs.~2(c) and 2(d) and the data obtained from another
independent sample studied using HB1A.

\begin{figure}[thb!]
    \includegraphics[width=3.2in]{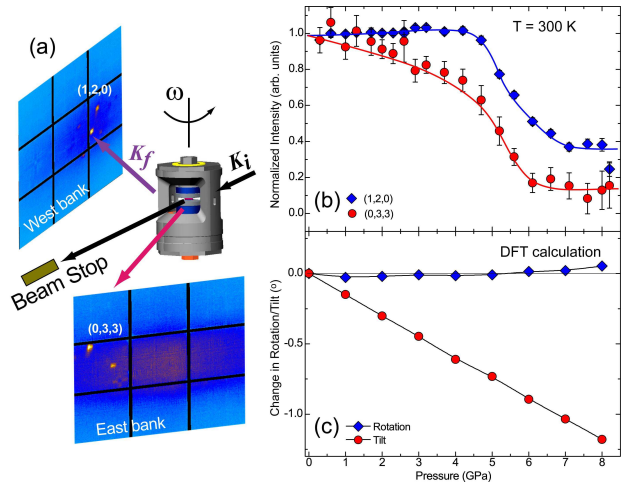}
    \caption{ (a) The pressure experiment setup with the crystal oriented such
    that the desired reflections can be measured simultaneously using both
    detector banks. (b) The pressure dependence of $I(P)/I(0)$ of the (1,2,0)
    and (0,3,3) reflections. Curves are guides to the eye. (c) DFT result of
    the change of in-phase rotation and antiphase tilt as a function of pressure.
    The calculation is only performed with structures in the $A2_1am$ phase.}
\end{figure}

Questions arise as to whether the antiphase tilt is more susceptible to
perturbation and whether the observed feature is pertinent only to this
particular compound. We thus investigate how the structure responds to the
external pressure as it provides an effective tuning
parameter to explore the phase diagram.  An earlier study of CMO  revealed a
structural transition from tetragonal to orthorhombic phase near $P \sim
1.3$~GPa \cite{Zhu02}.  However, the structure in the low-$P$ regime does not
agree with the established orthorhombic structure which is likely due to the
low sensitivity of x rays to oxygen. To characterize the transition,
we performed single crystal neutron diffraction under pressure using the SNAP
diffractometer. Figure 4(b) plots the intensities of the (1,2,0) and (0,3,3)
normalized to the (0,2,2) nuclear Bragg peak that has minimal change across
the transition. The (1,2,0) peak intensity associated with octahedral rotation
remains nearly unchanged for $\rm P < 4$~GPa while the (0,3,3) intensity is
suppressed immediately once the pressure is applied.  Both peak intensities
decrease at higher pressure and eventually evolve into pressure-independent
values for $P$ greater than the critical pressure $P_c \approx 7$~GPa,
suggesting the system enters another structure.  The $P$-dependent study
further corroborates the softness of the $X^{-}_3$ tilt mode which responds more
dramatically to the stimuli.

The suppression of the antiphase tilt and robustness of the inphase rotation
upon increasing $T$ and $P$ in CMTO exhibits distinct differences compared with the
isostructural $\rm (Ca_{1-x}Sr_x)_3Ti_2O_7$, where the antiphase tilt survives against
chemical substitution \cite{Oh15}. Previous investigation of the octahedral
instability in perovskite oxides has generalized rules to describe the phase
transition. (1) The pressure-induced change in octahedral rotation and tilt
$dR/dP$ decreases with increasing Goldshmidt tolerance factor (defined as
$\tau=(r_A+r_{\rm O})/\sqrt{2}(r_B+r_{\rm O})$, where $r_A$, $r_B$, $r_{\rm
O}$ are radii of the $A$-, $B$ cations, and O anion)
\cite{Goldschmidt26,Samara75}. This has explained the pressure behavior in
orthorhombic $\rm CaSnO_3$ \cite{Zhao04}, $\rm CaTiO_3$ \cite{Guennou10a}, and
$\rm SrTiO_3$ \cite{Guennou10b}.  (2) Materials with similar $\tau$ show
increasing $dR/dP$ as they move from $A^{3+}B^{3+}{\rm O}^{2-}_3$ to
$A^{2+}B^{4+}{\rm O}^{2-}_3$, and finally to $A^{1+}B^{5+}{\rm O}^{2-}_3$,
which is attributed to the pressure-dependence of the bond-valence parameter.
When the $A$-site cation is at high formal charge, the $A{\rm O}_{12}$
polyhedron is harder to compress because of its already small $A$-O
distance \cite{Shannon76,Angel05}. Extending to the $A_3B_2{\rm O}_7$ series,
calculation reveals a systematic trend of the distortion amplitudes with
respect to $\tau$ \cite{Mulder13}. The $a^0a^0c^+$ rotation is suppressed,
while the $a^-a^-c^0$ tilt remains stable as $\tau$ increases.  Although  CMO
and CTO have similar tolerance factors and the same formal charge of the
$A$-site ions, the observation in CMTO does not follow the aforementioned
rules and is not consistent with the robustness of the $a^-a^-c^0$-type tilt
mode. Recently, the role of orbital hybridization between the $B$ ions and the
oxygen has been suggested to play an important role in determining the
octahedral distortion \cite{Xiang17}.  When pressure is applied on systems
containing $B$ ions with empty low-lying $d$ states, the next-nearest neighbor
$B$-O inter-atomic distance is reduced, and the hopping between the O-$2p$
and $B$-$3d$ orbitals is enhanced significantly, which ultimately leads to the
increase of tilt/rotation. In contrast, the partially occupied $d$ orbitals
(where there is electron occupation in any of the five $d$ orbitals) will
suppress the distortion \cite{Xiang17}. This mechanism has successfully
explained the contrasting behavior of $\rm CaTiO_3$ and $\rm CaMnO_3$ under
pressure (see Supplementary Material \cite{Ca327SM}).  We apply similar
density functional theory (DFT) calculation to $\rm Ca_3Mn_2O_7$  and the
results are shown in Fig.~4(c).  Upon increasing pressure, the antiphase tilt
is significantly suppressed and the inphase rotation of $\rm MnO_6$ shows
signs of slightly increasing.  The nice agreement between calculation and
experimental data reveals the partially occupied $d$ orbitals of the Mn sites
are indeed critical for the stability of the MnO$_6$ distortion. Furthermore,
the contradictory pressure behavior between in-phase rotation and antiphase
tilt hints at the competition between the two and a presence of trilinear
coupling involving both octahedral distortion and the antipolar distortion
mode \cite{Xiang17}.

\begin{figure}[thb!]
    \includegraphics[width=3.2in]{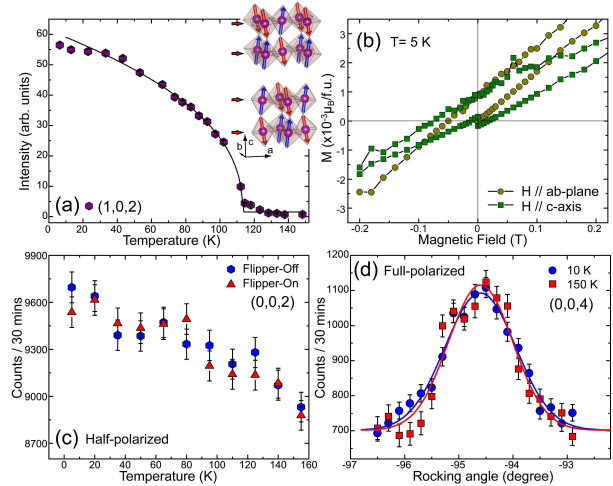}
    \caption{ (a) $T$ dependence of the (1,0,2) magnetic reflection from the
    $G$-type AFM order.  Inset shows the symmetry-allowed spin configuration. A
    canted magnetic structure leads to a ferromagnetic component along the $a$ axis.
    (b) Magnetization measurement with field perpendicular and parallel to the
    $ab$ plane. (c) $T$ dependence of the (0,0,2) peak with half-polarization setup.
    (d) Comparison of the scans across the (0,0,4) reflection at 10 and 150
    K with full-polarization setup. }
\end{figure}

CMTO enters antiferromagnetic (AFM) order at $T_N=115$~K and the
moments are dominantly along the $c$ axis [Fig.~5(a)]. The octahedral distortion is
directly coupled to the magnetic order and additional spin-orbit
interaction (SOI) gives rise to a canted moment of 0.06~$\mu_B$/site according to
the first principles calculation or $0.4 \pm 0.2~\mu_B$ from neutron powder
study \cite{Lobanov04}.  The symmetry-allowed weak ferromagnetism ($F$)
results from the trilinear coupling between the $G$-type AFM, the tilt
$X^{-}_3$, and $F$ in the lowest-order magnetoelastic coupling approximation.
The energy gain is of $V_{MX} \sim -G_c({\bf q}_1)F_bX^{-}_3({\bf q}_1)+G_c({\bf
q}_2)F_aX^{-}_3({\bf q}_2)$, where $a,b,c$ denote the crystallographic axes,
and ${\bf q}_1,{\bf q}_2$ are the ordering vectors (1/2,1/2,0) and
(1/2,-1/2,0) \cite{Harris11}. Since ferromagnetic scattering overlaps with
the nuclear scattering, polarized neutron measurement is used to
differentiate between the magnetic and nuclear contributions. The (0,0,2)/(0,0,4) peaks
are chosen for their large ratio of the magnetic over nuclear structure
factor. In the half-polarization configuration, the intensity is proportional
to the summation of nuclear and magnetic scattering in the flipper-off
channel, and becomes the difference between the two when the flipper is
turned on. The $T$ dependence of the (0,0,2) reflection in both
configurations shows no difference within standard error and no anomaly near
the AFM transition [Fig.~5(c)]. We also employed full polarization analysis
where the scattering intensity is dominantly magnetic in the spin-flip (SF)
configuration with polarization parallel to the moment transfer. In
Fig.~5(d), the scan profiles in the SF channel at $T=5$ and 150~K (above $T_N$) 
are essentially the same indicating the peak originates mainly
from the nuclear scattering leakage in the SF channel. Based on the
statistics of the (0,0,4) and other measured nuclear reflections, the
upper-bound of the ferromagnetic moment is determined to be $M_s=0.04~\mu_B/$Mn. 
Such a small value is somewhat unexpected when the
antisymmetric Dzyaloshinsky-Moriya interaction could lead to a sizable canting
moment \cite{Dzyaloshinsky58,Moriya60}, but consistent with
magnetization measurement [Fig.~5(b)], where the obtained moment is in the
order of $10^{-3} \mu_B$ per formula unit \cite{Jung00,Zhu12}. A reasonable
explanation is the weak SOI associated with the $3d$ Mn ion compared to
$5d$-electron iridates \cite{Ye13}. Although small, the unambiguous existence
of the canted moment in conjunction with the soft oxygen octahedral tilt mode
offers a feasible strategy to achieve magnetoelectricity control in this
layered perovskite, {i.e.,~}switching the direction of net moment with
electric field \cite{Oh15,Gao17}.

In summary, we performed a systematic neutron and x-ray diffraction study
characterizing the structural evolution of the hybrid improper multiferroic
CMTO.  We identified that the magnitude of the antiphase tilt of the MnO$_6$
octahedron decreases with increasing temperature and pressure, while the
inphase rotation remains stable. We attribute the suppression of antiphase
octahedral tilt to the partially occupied $d$ orbitals of the Mn$^{4+}$ ions,
in contrast with enhancement of both antiphase and in-phase distortion in CTO which
only has an empty $d$ orbital at the Ti$^{4+}$ site. The canted ferromagnetic
moment has an upper limit of $0.04~\mu_B$ despite substantial $\rm MnO_6$
tilt. Our study indicates that the active control of the octahedral
distortion provides a generic route for designing new ferroelectrics and
multifunctional materials.

\begin{acknowledgments}
Research at ORNL was sponsored by the Scientific User
    Facilities Division, Office of Basic Energy Sciences, U.S.~Department of
    Energy.  The work at Rutgers University was supported by the DOE under
    Grant No.~DE-FG02-07ER46382.  J.C.W. and J.M.S. acknowledge support from
    China Scholarship Council. This work has been partially supported by
    UT-Battelle, LLC under Contract No.~DE-AC05-00OR22725 for the
    U.S.~Department of Energy.
\end{acknowledgments}


\begin{thebibliography}{10}
\bibitem{Ca327twinning}
The structural refinement determines that single crystal used in the
  measurement has one twinned domains rotating 90 degree along the $c$-axis,
  the volume faction ratio between the primanry and twinned domain is 55:45.
\end{thebibliography}
\end{document}